\documentclass[useAMS,usenatbib]{mn2e}

\usepackage{times}
\usepackage{graphics,epsf}
\usepackage{amsmath}                
\usepackage{amsfonts}               
\usepackage{amssymb}                
\usepackage{epsfig}                 
\usepackage{rotating}
\usepackage{color}
\usepackage{multirow}
\usepackage{natbib}

\usepackage{epstopdf}

\title[Cosmology with SLSNe]{Cosmology with Superluminous Supernovae}
\author[D. Scovacricchi, R. C. Nichol, D. Bacon, M. Sullivan, S. Prajs]{D. Scovacricchi$^{1}$\thanks{E-mail: dario.scovacricchi@port.ac.uk}, R. C. Nichol$^{1}$, D. Bacon$^{1}$, 
M. Sullivan$^{2}$, S. Prajs$^{2}$\\
$^1$ Institute of Cosmology and Gravitation, University of Portsmouth, Portsmouth, UK\\
$^2$ School of Physics and Astronomy, University of Southampton, Southampton, SO17 1BJ, UK}

\begin{document}

\maketitle

\label{firstpage}

\begin{abstract}
We predict cosmological constraints for forthcoming surveys using Superluminous Supernovae (SLSNe) as standardisable candles. Due to their high peak luminosity, these events can be observed to high redshift ($z\sim 3$), opening up new possibilities to probe the Universe in the deceleration epoch. We describe our methodology for creating mock Hubble diagrams for the Dark Energy Survey (DES), the ``Search Using DECam for Superluminous Supernovae" (SUDSS) and a sample of SLSNe possible from the Large Synoptic Survey Telescope (LSST), exploring a range of standardisation values for SLSNe. We include uncertainties due to gravitational lensing and marginalise over possible uncertainties in the magnitude scale of the observations (e.g. uncertain absolute peak magnitude, calibration errors). We find that the addition of only $\simeq100$ SLSNe from SUDSS to 3800 Type Ia Supernovae (SNe Ia) from DES can improve the constraints on $w$ and $\Omega_{m}$ by at least 20\% (assuming a flat $w$CDM universe). Moreover, the combination of DES SNe Ia and 10,000 LSST-like SLSNe can measure $\Omega_{m}$ and $w$ to 2\% and 4\% respectively. The real power of SLSNe becomes evident when we consider possible temporal variations in $w(a)$, giving possible uncertainties of only 2\%, 5\% and 14\% on $\Omega_{m}$, $w_0$ and $w_a$ respectively, from the combination of DES SNe Ia, LSST-like SLSNe and Planck. These errors are competitive with predicted Euclid constraints, indicating a future role for SLSNe for probing the high redshift Universe.
\end{abstract}


\section{Introduction}
Type Ia Supernovae (SNe Ia) are vital standard candles in cosmology,
providing compelling evidence for the accelerated expansion rate of
the Universe \citep{riess98,perl99}, possibly caused by Dark Energy
\citep[see][as review]{amendola_book}. In the near future, observations of SNe Ia will be limited to relatively low
redshifts ($z\simeq1$), apart from
a handful of events studied by the Hubble Space Telescope \citep[e.g.][]{garnavich1997}. This is due to the low intrinsic
ultraviolet flux of Type Ia supernovae (redshifted into the infra-red at $z>1$), and the
lack of large-area space-based infra-red searches, e.g.
\citet{astier2014}.

Therefore one of the challenges of present day cosmology is
discovering new classes of high redshift standard candles \citep[see][]{king2014} to help break key degeneracies between cosmological
parameters and study the Universe far into the deceleration phase of its
expansion history. So far, several high redshift standard candle
candidates have been proposed, e.g. active galactic nuclei 
\citep{watson2011} and gamma ray bursts \citep{ghirlanda2006}, but
none have yet reached the level of standardisation currently achieved
with SNe Ia \citep{betoule2014}.

In recent years, the extensive search for SNe has led to the discovery
of a new class of SN explosion, some up to a hundred times brighter
than normal SNe Ia and core collapse SNe. These new SNe, named
`Superluminous Supernovae' (hereafter SLSNe), have exceptional peak
magnitudes ($M_{U}\lesssim-21$ magnitudes) and are characterized by a total radiated
energy of $\sim10^{51}$ erg; see \citet{Gal-Yam2012} for a review. To
date, only tens of SLSNe have been detected and studied, but already
the SLSN population shows some diversity and has been classified into
possible sub-classes based on photometric and spectroscopic
properties: Type I SLSNe are spectroscopically-classified as hydrogen
free, while Type II SLSNe show some hydrogen emission lines possibly
due to interactions with circumstellar material (CSM).
Finally, Type R SLSNe are characterized by long,
slowly-declining light curves and are possibly pair--instability
SNe \citep{Gal-Yam2012}, although \citet{nicholl2013} have suggested that
the Type I/R SLSN class should be re-classified as Type Ic SLSNe
(SLSNe Ic), analogous to the normal Type Ic SNe.

Recently, SLSNe-Ic have been proposed as a new standardisable candle in
cosmology despite uncertainties about the physical nature
and categorization of these transient events. This is driven by the
fact that SLSNe Ic show less dispersion in their bolometric light
curves (e.g. $\simeq 0.25$\,mag after 25 rest-frame days; see \citealt{Papadopoulos2015}) and are several magnitudes brighter at peak
than SNe Ia, meaning they can be discovered to high redshift; e.g.
\citet{cooke2012} have discovered a candidate SLSN at $z=3.90$.

\cite{inserra2014} studied a sample of 16 SLSNe Ic, over the redshift
range $0.1<z<1.2$ (4000 \AA{} to 20000 \AA{} rest-frame), to develop a method for standardizing the light
curves of SLSNe. In their analysis, they found a simple relationship
between the peak magnitude and the colour and decline rate of the
SLSNe Ic measured in two synthetic filters (chosen to reduce the
effects of spectral features). This standardisation displayed a
scatter (r.m.s.) in the corrected magnitudes of between 0.19 and
0.26 mag for the rest frame color evolution, with the exact value depending on the methodology and the details of the SLSN Ic
sample used (this range of scatter is consistent with that seen by \citealt{Papadopoulos2015}). This study suggests SLSNe Ic could be standardized to an
accuracy approaching normal SNe Ia (typical r.m.s. of 0.14\,mag in
corrected peak magnitude), and could be competitive with SNe Ia with better
modelling of a larger sample of SLSNe light curves. 

We study here the prospect for cosmological constraints using SLSNe as
standard candles. Recently, \citet{wei2015} performed a similar
analysis but focused on the power of existing SLSNe Ic, from
\citet{inserra2014}, to differentiate between competing cosmological
models, namely Lambda Cold Dark Matter ($\Lambda$CDM) and the
$R_{\rm h}=ct$ model. They concluded that present data was insufficient
to differentiate between these models, but showed that samples of
several hundreds of SLSNe Ic events would be sufficient, demonstrating the
possible constraining power of these high-redshift objects.

In this paper, we focus on the possible parameter constraints for the
concordance $\Lambda$CDM cosmology together with $w$CDM cosmologies
including an evolving dark energy equation-of-state with $w
=w(a)$, a linear function of the scale factor $a$. We create a series of mock Hubble Diagrams (HDs) for a set
of realistic SNe Ia and SLSNe Ic samples, and find confidence limits
on cosmological parameters from Markov Chain Monte Carlo (MCMC) fits to these HDs. We also investigate
the impact of a range of standardisation values
\citep{inserra2014,Papadopoulos2015} and consider the effect of
lensing magnification, which will be critical for such high redshift
events \citep{marra2013}.

In Section 2, we present details of new searches for SLSNe. In Section 3, we provide the methodology used to create our mock Hubble Diagrams for a number of on-going and planned surveys. Section 4 outlines how we then analyse those mock data with cosmological fitting discussed in Section 5. We present results on Section 6 and conclude in Section 7. We assume throughout a flat $\Lambda$CDM Universe with
$\Omega_m=0.3$ and $H_0=68$ km\,s$^{-1}$\,Mpc$^{-1}$ as our 
fiducial cosmology, consistent with Planck \citep{planck_collaboration}.

\begin{figure*}
\includegraphics[width=.6\textwidth]{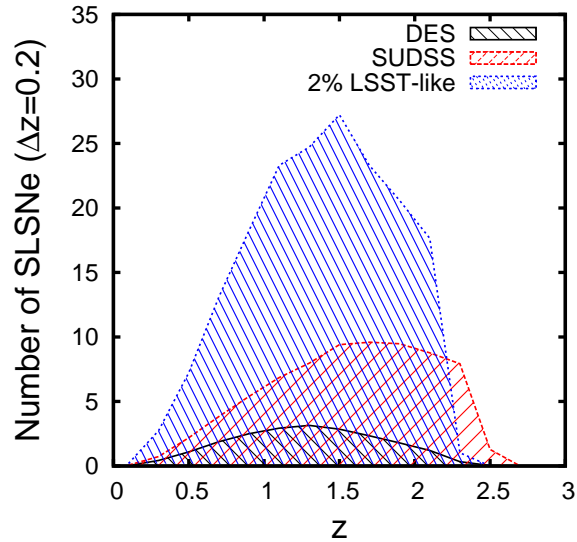}
\caption{The redshift histogram for well-observed SLSNe-Ic from SUDSS and DES (each assuming a 3-year survey) and LSST-like (assuming 2\% of the total number of SLSNe detected in 10 years). We consider well-observed supernova light curves to have $>$5 detectable epochs, in at least two filters each, a pre-explosion detection limit, and at least one supernova detection at $>$20 days past peak brightness.}
\label{sudss_zhist}
\end{figure*}

\section{Search for Superluminous Supernovae}
\label{chapter_sudss}

Over the next decade, several on-going and planned experiments will
increase the number of known SLSNe by at least an order of magnitude,
making it possible to use these objects to probe cosmology. For
example, the Dark Energy Survey (DES) supernova programme \citep{bernstein2012} has already detected several SLSNe in its first season of data 
\citep{Papadopoulos2015} with the potential for detecting
further events over the next few years.  Moreover, the well-sampled
DES multicolour $griz$ light curves of these new SLSNe are better than those
presently available in the literature, which will further help in the
study of SLSN standardisation.

Unfortunately, the observing strategy of DES is not optimal for
characterising all the SLSNe it will detect due to temporal edge
effects.  At high redshift, time dilation extends the light curve
duration of SLSNe to many months in the observer frame, leading to
incomplete data if the start, or end, of the light curve extends
outside the DES observing season. To alleviate this problem, we have
begun the `Search Using DECam for Superluminous Supernovae' (SUDSS; PI: Sullivan).
This program supplements the DES SN programme with the addition of extra
epochs either side of the nominal DES season (September to early
February) extending the observing window to nearly eight months, thus
mitigating some of these temporal edge effects. 

SUDSS has also begun
long-term monitoring of several non-DES fields, namely 6 ${\rm
  deg^2}$ (two pointings of the Dark Energy Camera, or DECam,
\citealt{flaugher2015}) near the HST Cosmic Evolution Survey 
\citep[COSMOS:][]{scoville2007} field, and 6 ${\rm deg^2}$ in a new field close
to the South Ecliptic Pole to facilitate near year-long monitoring.
When combined with DES fields, SUDSS observes approximately 23 ${\rm
  deg^2}$ in \emph{griz} with an average cadence of 14 days in the
observer-frame, which is optimal for tracking such long-duration light
curves.  SUDSS began collecting data in 2014 and is planned for a
three-year survey.

In Fig.~\ref{sudss_zhist}, we present the predicted SLSN Ic redshift
histogram for SUDSS and DES. This prediction was made using a simple
Monte Carlo simulation of the SUDSS (and DES) surveys. We simulated
SLSNe using the Spectral Energy Distribution (SED) model of
Prajs et al. (in prep.), which is based on a series of magnetar model fits
to SNLS-06D4eu \citep[][]{howell2013} together with a $k$-correction
spectral template. The volumetric SLSN rates of \citet{quimby2012} and
\citet{cooke2012} were used to give the number of SLSNe exploding in
the SUDSS and DES search volumes as a function of redshift during the
observing seasons of both surveys.  

We then calculated the supernova
magnitudes, as a function of time, using the cadence of each survey,
the redshifted SLSN SED model, and assuming our fiducial cosmology (Section 1). The predicted magnitudes were then compared to
the DECam exposure time calculator spreadsheet (using v6 from March 2015 available on the DECam CTIO
webpage) to assess the detectability of the SNe given the magnitude
limits of both surveys. For DES, we use the magnitude limits in $griz$ from \citet{bernstein2012}, while SUDSS is expected to reach a depth (per epoch) of 
24.6, 24.5, 24.4 and 24.1 in $griz$
respectively (AB magnitudes for a 5-sigma point source). The depth of SUDSS is approximately half a magnitude deeper than the DES shallow fields in the redder bands. 

We emphasise that these predictions are only indicative, as there are
uncertainties in the rate of SLSNe Ic (especially with redshift),
their spectra and light curve evolution, and luminosity function.
However, SUDSS should discover $\simeq$75 high-quality SLSNe-Ic (see
Fig.~\ref{sudss_zhist}),
over the redshift range $0.1<z<2.5$, assuming the full three years of
observations, each with a well-sampled light curve. SUDSS will also find many Type II SLSNe. Such
samples will allow us to characterise the luminosity functions of
these different sub-classes of SLSNe, while greatly improving their
possible standardization \citep{inserra2014} as cosmological probes.
We will also have sufficient SLSNe to differentiate between competing
cosmological models \citep[e.g.][]{wei2015} and improve parameter constraints
for the concordance $\Lambda$CDM model.

In comparison, Fig.~\ref{sudss_zhist} also shows the expected SLSN Ic redshift distribution for DES alone, which has a
predicted total of just $\simeq$15--20 SLSNe over three years. SUDSS improves considerably on this due to the likely higher SLSN rate at
$z>1$ \citep{cooke2012}, where SUDSS is more sensitive, and the
decrease in temporal edge effects because of the longer observing
window.

In Fig.~\ref{sudss_zhist}, we also show our prediction for a sample of SLSNe that maybe possible from the Large Synoptic Survey Telescope (LSST), which we call LSST-like from hereon. We use the same methodology as used for SUDSS, but adjust the depth to that expected for the LSST wide survey, namely 25.0, 24.7, 24.0, 23.3 in $griz$ respectively (AB magnitudes for a 5-sigma point source; \citealt{lsst_science_book}). This depth is slightly less than SUDSS, especially in the $i$ and $z$ bands, which is seen in the drop in sensitivity of SLSNe at $z>2$ in Fig.~\ref{sudss_zhist} compared to SUDSS. In the simulations, we have also accounted for the nominal LSST wide survey cadence of one filter every three days, compared to DES and SUDSS which routinely observing all filters per field visit, but at a lower cadence per field. 

\section{Mock catalogues}
\label{chapter_mock}

We describe here the procedure used to generate mock Hubble Diagrams
for the three samples of supernovae of interest within this paper.
These are a DES-like sample of SNe Ia, based on the predictions of
\citet{bernstein2012}, a SUDSS SLSNe sample as represented by Figure
\ref{sudss_zhist}, and a possible LSST-like sample. Throughout this paper, we name these three samples as DES, SUDSS and
LSST-like respectively.


For each mock supernova in each of our three samples (SNe Ia or SLSNe), we begin by calculating their cosmological distance modulus ($\mu_{\rm cos}$) assuming a Hubble rate ($H(z)$) for a flat $\Lambda$CDM cosmology,
\begin{equation}
H(z)=H_0\left[\Omega_m \left(1+z \right)^3 + \left(1-\Omega_m \right) \right]^{\frac{1}{2}},
\end{equation}
with $H_0$ and $\Omega_m$, the Hubble Constant and the matter density parameter at the present epoch respectively (note that the exact choice of $H_0$ is irrelevant, see Section \ref{chapter_analysis}). Where appropriate, we assume our fiducial cosmology given in Section 1. 

\subsection{Errors in the Distance Modulus}
\label{distance}

Two sources of uncertainty are present in $\mu_{\rm obs}$ (the
distance modulus recovered from observations) that make it different
from $\mu_{\rm cos}$.  The first is $\delta \mu_{\rm err}$, a
combination of several statistical uncertainties, including
measurement error. This adds scatter to the Hubble diagram, and it is
well-known that part of this scatter is not accounted for in the
experimental error budgets of SN surveys, resulting in larger than expected $\chi^2$ values when
fitting cosmological models. This is usually accounted for by adding
an `intrinsic scatter' to the whole SN population ($\sigma_{int}$) to
obtain acceptable $\chi^2$ values, with the value of $\sigma_{int}$
potentially varying between different surveys \citep[][]{conley2011}. We stress our $\delta \mu_{\rm err}$ mimics the effect of
$\sigma_{int}$ in our analysis but is more broadly defined to include all potential sources of statistical error in SN distance moduli. We assume the average value of the error is zero, i.e., $\langle \delta \mu_{\rm err} \rangle =0$, as we deal with systematic offset in the magnitude systems of SNe in section
\ref{chapter_analysis}.

The second uncertainty is the effect of gravitational lensing along
the line of sight to each SN ($\delta \mu_{\rm len}$), which can add
further scatter to the Hubble diagram, especially for the higher
redshift SLSNe. Again, we note that the mean lensing magnification of objects is zero across the sky, i.e. $\langle \delta \mu_{\rm len} \rangle = 0$.

To compute $\delta\mu_{\rm len}$, we begin with the definition of the
convergence $\kappa$ as an integral along the line of sight, written
for a lensed source located at the comoving distance
$\chi_i=\chi(z_i)$ (see \citealt{bartelmann2001,schneider2005} for the
following equations)
\begin{equation}
\label{convergence_def}
 \kappa \left(\overrightarrow{\theta},\chi_i \right)=\frac{3H_0^2\Omega_m}{2c^2}\int_0^{\chi_i}d\chi \frac{\chi (\chi_i - \chi)}{a\left( \chi\right) \chi_i}\delta\left(\chi \overrightarrow{\theta},\chi \right).
\end{equation}
A measure of the lensing variance is the convergence \emph{rms} on a line of sight, with a pre-factor converting this into its effect on magnitude:
\begin{equation}
\label{sigma_len}
 \sigma_{\rm len}^2 =\left[ \frac{5}{\ln(10)} \right]^2 \frac{1}{2\pi}\int_0^{\infty} lP_{\kappa}(l)dl,
\end{equation}
where $P_{\kappa}(l)$ is the convergence power spectrum, as a function of the angular wavenumber $l$ of,
\begin{equation}
\label{P_k}
 P_{\kappa}(l)=\frac{9H_0^4\Omega^2_m}{4c^4}\int_0^{\chi_i} d\chi \frac{\left(1-\frac{\chi}{\chi_i} \right)^2}{a^2\left(\chi \right)}P_{\delta}\left(\frac{l}{\chi},\chi \right).
\end{equation}
In Equations~\ref{convergence_def} and \ref{P_k}, $c$ is the speed of
light, $\overrightarrow{\theta}$ is the initial direction of the light
propagation, $\delta$ is the density contrast, $a=a\left(\chi \right)$
is the scale factor and $P_{\delta}\left(\frac{l}{\chi},\chi \right)$
the total matter power spectrum, a function of the Fourier mode
$k=l/\chi$ and time (via $\chi=\chi(t)$). Equations \ref{sigma_len}
and \ref{P_k} show that $\sigma_{\rm len}$ is a function of the
cosmological model and that it depends on the redshift of the lensed
source $z_i$ \citep{bartelmann2001}. In Fig.~\ref{int_vs_len}, we show
our prediction for $\sigma_{\rm len}(z_i)$ as a function of redshift
and, as expected, it grows monotonically with $z_i$ as the light path
becomes longer.

We treat here gravitational lensing as an additional source of noise, but we note this effect could be used in the future to provide additional cosmological constraints \citep[see][]{quartin2014} especially if the skewness and kurtosis of the SN magnification distribution can be measured as a function of redshift. Moreover, the lensing effect could be measured through cross-correlation with foreground structures (see \citealt{jonsson2010,smith2014}) thus improving the scatter on the Hubble diagram as the effects of lensing could be estimated and decreased. We do not discuss these lensing signals further in this paper but note they will become important for DES and LSST in the future.

\subsection{Constructing Samples}
\label{constructing_samples}

We now create mock Hubble diagrams for each of our SN samples (DES, SUDSS, LSST-like). First, we randomly draw
a SN redshift from the redshift distribution appropriate for each
survey until we have obtained the expected total number of
events for the survey in question. For DES, we assume the redshift distribution for the {\it hybrid 10} simulation shown in \citet{bernstein2012}, which is a
combination of two `deep' and eight `wide' DES fields, and consistent
with the on-going DES SN strategy. This provides a total of 3500 SNe Ia for the final DES sample to which we add a further 300 $z<0.1$ SNe Ia to reflect the expected number of high-quality, local SNe Ia available in forthcoming surveys. 

For SUDSS, we use the redshift histogram as given in Fig.~\ref{sudss_zhist} which has a total of 73 SLSNe Ic. We also add a further 25 $z<0.3$ SLSNe Ic to this distribution to simulate the likely low-redshift sample from surveys like Skymapper \citep{skymapper}, PTF \citep{ptf} and PanSTARRS \citep{panSTARRS} that are routinely finding a few high-quality, local SLSNe per year. In both cases (DES and SUDSS), we assume these local SNe possess the same characteristics as the higher redshift objects, i.e., there is no evolution in their properties or systematic differences in their photometry.  

For our LSST-like sample, we use the histogram as given in Fig.~\ref{sudss_zhist}, assuming a total number of 10,000 SLSNe detected over the 10-year operations of the LSST, i.e., a factor of $\sim$100 larger than SUDSS. This larger number reflects the greater volume probed by the 18,000 deg$^2$ LSST wide survey, and is consistent with the predicted rate of luminous supernovae as given in the LSST Science Book (Table 8.3 in Chapter 8; \citealt{lsst_science_book}). Further simulations are required to obtain a more accurate prediction for the total number of LSST SLSNe.    

\begin{figure*}
\includegraphics[width=0.5\textwidth]{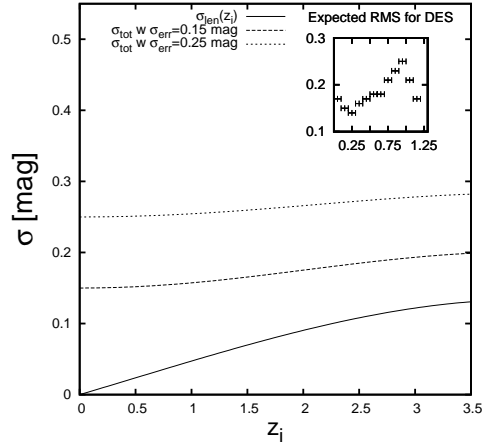}
\caption{$\sigma_{\rm tot}$ as a function of the source redshift $z_i$ for different values of $\sigma_{\rm err}$ (see caption). The solid line is the contribution from lensing (Equation \ref{sigma_len}) which is added in quadrature to $\sigma_{\rm err}$ in our analysis. \emph{Inset}: The predicted root--mean--square magnitude error for DES as a function of redshift taken from Bernstein et al. 2012 (the error bars along the x-axis show the bin size of $\Delta z=0.1$).}
\label{int_vs_len}
\end{figure*}

Next, given the redshift of each SN ($z_i$), we calculate the cosmological distance modulus ($\mu_{\rm cos}$, Section \ref{distance}) with an additional  error $\delta\mu_{\rm err}$ for each SN drawn at random from a Gaussian of fixed width of $\sigma_{\rm err}$ which can vary for each survey. We also add the effect of lensing for each supernova ($\delta\mu_{\rm len}(z_i)$) by drawing a random value from a Gaussian distribution with zero mean and a standard deviation given by Equation \ref{sigma_len}. Therefore, the value of $\mu_{\rm obs}$ is simulated as the sum of all three quantities \citep{dodelson2005}, 

\begin{equation}
\label{delta_mu}
 \mu_{\rm obs}(z_i)=\mu_{\rm cos}(z_i)+\delta\mu_{\rm err}+\delta\mu_{\rm len}(z_i)
\end{equation}

\noindent where $z_i$ is the redshift of each SN in the sample. 

For each supernova, the error bar on the distance modulus is then $\sigma_{\rm tot}=\sqrt{\sigma^2_{\rm err}+\sigma^2_{\rm len}}$.
In Figure \ref{int_vs_len}, we show the comparison of $\sigma_{\rm tot}$ for different values of $\sigma_{\rm err}$ and the expected lensing scatter\footnote{The non-linear corrections to the matter power spectrum have been computed following the approach of \cite{smith2003}, starting from a 
linear power spectrum for adiabatic CDM with transfer function by \cite{eisenstein1999}. The approximated growth factor used is from \cite{carroll1992}. The clustering parameter $\sigma_8$ is set to the value 0.79.} using Equation \ref{sigma_len}. The lensing significantly contributes for the lower value of $\sigma_{\rm err}$ at $z > 2.5$, so it is important to include this noise term for our SLSNe samples. Increasing the value of $\sigma_{\rm err}$ decreases the importance of the lensing dispersion over the redshift range studied here.

For DES, we assume the $\sigma_{\rm
  err}$ is the product of two errors added in quadrature. First, we include the r.m.s. reported in Table 14 of \citet{bernstein2012}, and given in Fig.~\ref{int_vs_len} for completeness. We allow this to vary across redshift as shown. Second, we include an additional term $\delta \mu_{\rm sys}$ to take into account the possible effect of systematics. As before, this random number is distributed accordingly to a Gaussian distribution, with zero mean and standard deviation $\sigma_{\rm sys}$ selected to be 0.1 mag in order to reproduce the contour widths showed in figure 29 of \citealt{bernstein2012} (see Section 5 for detail). In this case, the error on the distance modulus is given by $\sigma_{\rm tot}=\sqrt{\sigma^2_{\rm err}+\sigma^2_{\rm len}+\sigma^2_{\rm sys}}$.

\begin{figure*}
\includegraphics[width=1.1\textwidth]{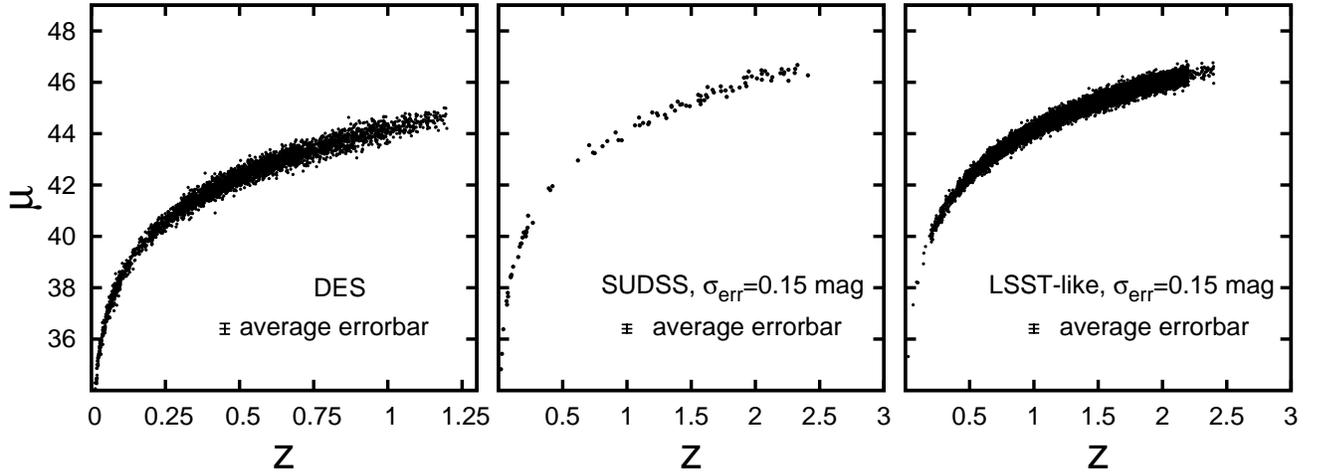}
\caption{The mock Hubble diagrams generated for DES, SUDSS and LSST-like (from left to right). The last two have $\sigma_{\rm err}=0.15$ magnitudes (for the case shown in this plot).}
\label{HD_multiplot}
\end{figure*}

For the SUDSS and LSST-like sample, we must assume a value for $\sigma_{\rm err}$
(constant across all SNe) and choose two values, namely
$0.15$ and $0.25$ magnitudes to reflect the range of uncertainty in
the possible standardisation of SLSNe discussed in \citet{inserra2014}. 
In Figure \ref{HD_multiplot}, we show the mock HDs used in this analysis for the DES, SUDSS and LSST-like surveys (the latter two using $\sigma_{\rm err}=0.15$ mag in this instance).

\section{Analysis of the Hubble Diagram}
\label{chapter_analysis}

We now fit our mock Hubble diagrams to determine the cosmological parameter constraints we can achieve for forthcoming surveys. When fitting only supernova data (both SLSNe and SNe Ia), we use a custom likelihood code based on the Downhill method of maximisation (Numerical Recipes by \citealt{press1992}) while the marginalization over the cosmological parameters (nominally $w$ or $\Omega_m$ in the $w$CDM model) is performed numerically by using the Gauss-Legendre method from the GNU Scientific Library (but implemented in {\tt Fortran90\footnote{http://www.lrz.de/services/software/mathematik/gsl/fortran/}}). 

We employ a full MCMC code when we add information from the Cosmic Microwave Background (CMB) using Planck data \citep{planck_collaboration}. For this, we use a modified version of the {\tt Fortran90} module provided for the Union 2.1 sample \citep{union21}, originally based on the module for supernovae released with the first version of CosmoMC. 

In both cases, if we did not consider magnitude calibration issues, we could use the following likelihood, obtained by considering the distance modulus measurements as independent, and all the errors as Gaussian,
\begin{equation}
  \label{likelihood_eq}
 \mathcal{L}=\frac{1}{(2\pi)^{n/2}\sqrt{\det C}}\exp \left[-\frac{1}{2} \left(\overrightarrow{\Delta \mu}^T C^{-1} \overrightarrow{\Delta \mu}\right)\right],
\end{equation}
where $\overrightarrow{\Delta \mu}=\overrightarrow{\mu}_{\rm obs}-\overrightarrow{\mu}_{\rm cos}$ is the $n$-dimensional data vector and $n$ is the number of supernovae in the sample.
Neglecting the covariance between data (i.e. all the non-diagonal terms are set to be zero), the covariance matrix would simply be given by
\begin{equation}
 C_{ij}=\langle\Delta\mu_i \Delta\mu_j \rangle = \sigma^2 _{ij}\delta_{ij}=\sigma_{\rm err}^2+\sigma_{\rm len}^2(z_i)
\end{equation}
where each measurement has an error equal to the sum in quadrature of the data and lensing uncertainties discussed in Section \ref{chapter_mock}. 

However in reality, when fitting the mock HDs we must include the possibility of an unknown, overall offset in magnitude $\xi$. In our analysis, this nuisance parameter accounts for the contribution of the Hubble Constant to the distance modulus (i.e. $5 \log \frac{100}{H_0}$) and, when fitting real SN data, this normalization parameter also accounts for the unknown value of the mean (corrected) absolute magnitude ($M$) of the SN population as well as any photometric calibration offset. Since we simulate, and fit directly, the distance moduli, we do not assume any value for $M$, and then $\Delta \mu_i$ becomes
\begin{equation}
\Delta \mu_i=\mu_{obs,i}-\mu_{cos,i}(h=1)+\xi.
\end{equation}

To marginalize over $\xi$, we follow the analytical procedure given in \cite{press1993} and \cite{bridle2002} which considers the marginalization of an unknown calibration uncertainty with a flat prior, under the hypothesis that the likelihood is Gaussian. In this particular case, the marginalization can be computed analytically, and the marginalized likelihood $\mathcal{L}_{\xi}(\overrightarrow{\alpha}) \propto \int d\xi \mathcal{L}(\xi,\overrightarrow{\alpha})$, where $\overrightarrow{\alpha}$ is the array of cosmological parameters, is then
\begin{eqnarray}
-2\ln \mathcal{L}_{\xi}= \overrightarrow{d}^T \left(C^{-1} - \frac{C^{-1} \overrightarrow{v}\overrightarrow{v}^T C^{-1}}{\overrightarrow{v}^T C^{-1}\overrightarrow{v}} \right) \overrightarrow{d}+ \nonumber \\* 
+ \ln \left(\overrightarrow{v}^T C^{-1} \overrightarrow{v} \right) + {\rm const.}
\end{eqnarray}
where $\overrightarrow{v}$ is the $n$-dimensional vector of unitary components and $\overrightarrow{d}=\left(\overrightarrow{\mu}_{\rm obs}-\overrightarrow{\mu}_{\rm cos} \right)$. It follows that the $\mathcal{L}_{\xi}$ can be re-written in terms of an effective chi-squared $\chi^2_{\rm eff}$, such that
\begin{equation}
 \mathcal{L}_{\xi} \propto \exp{\left( -\frac{\chi^2_{\rm eff}}{2}\right)},
 \label{eq:Lxi}
\end{equation}
defining 
\begin{equation}
 \chi^2_{\rm eff}=\overrightarrow{d}^T \left(C^{-1} - \frac{C^{-1} \overrightarrow{v}\overrightarrow{v}^T C^{-1}}{\overrightarrow{v}^T C^{-1}\overrightarrow{v}} \right) \overrightarrow{d}.
 \label{eq:chi2l}
\end{equation}
We will therefore directly calculate the likelihood marginalized with respect to $\xi$ from Equations (\ref{eq:Lxi}) and (\ref{eq:chi2l}). In this way we do not need to fix a value for $\xi$ at the level of the mock HDs, and do not need to explore the $\xi$ direction of the parameter space (saving computational time) or fix its prior, which contributes only to the integration constant (and so does not affect the maximum likelihood procedure).
This method is included in CosmoMC (see Appendix F of \citealt{cosmomc} for detail). 

When combining SLSNe data with lower redshift SNe Ia, the likelihood is given by the product of two likelihoods,
\begin{equation}
 \mathcal{L}=\mathcal{L}_{\rm SNIa}*\mathcal{L}_{\rm SLSNe},
\end{equation}
where each likelihood has been marginalised over an unknown normalisation parameter (i.e., a $\xi$ for each sample) which is computed separately for SNe Ia and SLSNe using Equation (\ref{eq:chi2l}).

We use a combination of two methodologies when fitting for cosmological parameters. We use our own maximum likelihood code (called ``Lik" in this paper) described above when fitting for a flat cosmology with a Dark Energy fluid ($p_{\Lambda}=w\rho_{\Lambda}c^2$) assuming a constant equation-of-state parameter $w$ ($w$CDM). In this case, the degrees of freedom are $\Omega_m$ and $w$. When combining the supernova data with the CMB, we use CosmoMC fitting (``MCMC") instead for a flat $w$CDM model as well as a flat Universe with a linearly growing equation-of-state, namely $w(a)=w_0+w_a(1-a)$ \citep{chevallier2001}, or the $w_z$CDM model. The jointly fitted parameters are  $\Omega_{c}h^2$, $w_0$, $w_a$, (respectively the reduced dark matter density parameter and the two dark energy parameters) and $\log A$ (logarithmic fluctuation amplitude with pivoting scale 0.05 Mpc$^{-1}$). Other CosmoMC parameters\footnote{$\Omega_b h^2=0.0222$ (the reduced baryonic matter parameter), $\theta=1.0411$ (100 times the ratio of the angular diameter distance to the LSS sound horizon), $\tau=0.0925$ (optical depth at the reionization), $\sum m_\nu=0.06$ (sum of physical masses of standard neutrinos, with no sterile neutrino), $\Omega_K=0$ (curvature parameter), $n_{\rm run}=n_{\rm run,run}=0$ (running of the spectral index, running of the running of the spectral index), $r=0$ (ratio of tensor to scalar primordial amplitudes at pivot scale), $N_{\rm eff}=3.046$ (effective number of neutrinos), $\alpha^{-1}=0$ (correlated CDM isocurvature), $\Delta z_{\rm re}=0.5$ (width of reionization), $A_{\rm lens}=1$ (lensing potential scaled by $\sqrt{A_{\rm lens}}$), $f_{dm}=0$ (CosmoRec dark matter annihilation parameter), $n_s=0.96$ (spectral index), $A^{\phi\phi}_L=1$ (scaling of lensing potential power) } are set to their default values (using the April 2014 version of the code).

To test our methodology, we used our Lik code to fit for a flat $\Lambda$CDM model with one degree of freedom ($\Omega_m$) to both the published JLA data sample \citep{betoule2014} and a mock JLA Hubble diagram (constructed as discussed in Section \ref{chapter_mock}), with $\sigma_{\rm err}=0.17$ magnitudes (same as the intrinsic scatter of that sample), and the same overall number of SNe and redshift distribution. For each mock supernova, we assign an error bar that is the average error found by \citet{betoule2014}, namely 0.19 magnitudes. In both cases (real and mock data), we obtain $\Delta \Omega_m \simeq 0.018$ (at 68\% confidence limit) and thus conclude that our methods for creating and fitting our mock Hubble diagrams are realistic. The uncertainty on our measurement of $\Omega_m$ from the real JLA sample is fully compatible with the published result of \citet{betoule2014}, who find $\Omega_m=0.289^{+0.018}_{-0.018}$ without systematic errors.


\begin{table*}

\caption{Relative error (68\% confidence limit) for the cosmological parameters $\Omega_m$ and $w$ ($w$CDM model) using the likelihood code (\emph{Lik}) for different combinations of our mock sample data sets (DES, SUDSS and LSST-like) . For SUDSS and LSST-like, we provide results for $\sigma_{\rm err}=0.15$ and $0.25$ mag. For DES, we use the $\sigma_{\rm err}$ given in Figure 2 as derived from \citet{bernstein2012}.}

\begin{tabular}{ l | l | l | c | c }\hline \hline
Samples & No. of SNe  & $ \left|\frac{\Delta \Omega_m}{\Omega_m} \right|$ & $ \left|\frac{\Delta w}{w}\right|$ \\ \hline
DES &  3800 SNe Ia  & 0.11 & 0.11\\
\hline
$\sigma_{\rm err}=0.15$ mag\\
\hline
SUDSS & 73 SLSNe + 25 low-z SLSNe & 0.07 & 0.29 \\
LSST &  10000 SLSNe & 0.03 & 0.07\\
SUDSS+DES  & 73 SLSNe + 25 low-z SLSNe + 3800 SNe Ia  & 0.07 & 0.09 \\
LSST-like+DES & 10000 SLSNe + 3800 SNe Ia  &0.02 & 0.04\\

\hline
$\sigma_{\rm err}=0.25$ mag\\
\hline
SUDSS & 73 SLSNe + 25 low-z SLSNe & 0.14 & 0.34 \\
LSST &  10000 SLSNe & 0.03 &0.11 \\
SUDSS+DES  & 73 SLSNe + 25 low-z SLSNe + 3800 SNe Ia  & 0.11 & 0.10\\
LSST-like+DES  & 10000 SLSNe + 3800 SNe Ia  & 0.03 &0.04\\

\hline
\label{table_results}
\end{tabular}
\end{table*}

\begin{table*}

\caption{Relative error (68\% confidence limit) for the cosmological parameters $\Omega_m$, $w_0$ and $w_a$ (of which we only show the absolute error, since $w_a \simeq 0$), in a $w_z$CDM Universe, using  CosmoMC and different combinations of our mock data sets (DES, LSST-like and SUDSS). See text for details.
}

\begin{tabular}{l | l | c | c | c }
\hline\hline
Samples & No. of SNe & $\left|\frac{\Delta \Omega_m}{\Omega_m}\right|$ & $\left|\frac{\Delta w_0}{w_0}\right|$ & $\Delta w_a$\\ 
\hline
DES+Planck & 3800 SNe Ia & 0.030 & 0.090 & 0.366\\
\hline
$\sigma_{\rm err}=0.15$ mag\\
\hline
SUDSS+DES+Planck & 73 SLSNe + 25 low-z SLSNe + 3800 SNe Ia & 0.026 & 0.078 & 0.325\\
LSST-like+DES+Planck & 10000 SLSNe + 3800 SNe Ia & 0.016 & 0.045 &  0.143\\
\hline
$\sigma_{\rm err}=0.25$ mag\\
\hline
SUDSS+DES+Planck & 73 SLSNe + 25 low-z SLSNe + 3800 SNe Ia & 0.027 & 0.087 & 0.353 \\
LSST-like+DES+Planck & 10000 SLSNe + 3800 SNe Ia & 0.017 & 0.049 &  0.170\\
\hline
\end{tabular}
\label{table_results_w0wa}
\end{table*}

\section{Constraints on Cosmological Parameters}
\label{results}

In Table \ref{table_results}, we report our results for the $w$CDM model fitted to combinations of the three samples considered herein (DES, SUDSS, LSST-like). In this table, we do not quote the best fit values for the cosmological parameters, as these are all consistent (within 2$\sigma$) with the expected fiducial cosmological model assumed in the construction of the mock data. 

We instead quote statistical errors based on the width of likelihood functions as a detailed analysis of the possible systematic errors associated with these SN samples is beyond the scope of this paper. We have marginalised over an overall magnitude offset between the samples (Section 4), which should cover major systematic uncertainties in the photometric calibrations of the samples and the absolute magnitude scale. The  confidence limits on the cosmological parameters are computed by integration of the one dimensional posterior distributions, considering $68\%$ of the area around the mean (with no assumption on the shape of the posteriors). 

\begin{figure*}
\includegraphics[width=0.5\textwidth]{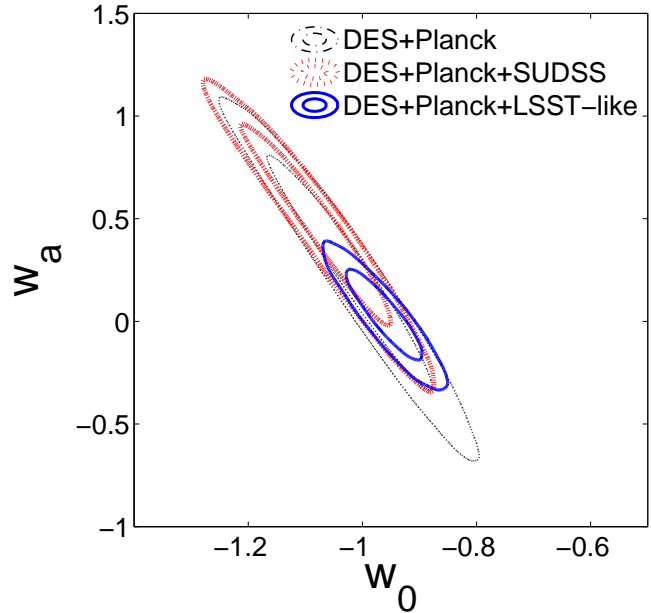}
\caption{The 68\%  and 95\% CL marginalized contours in the $w_0$-$w_a$ plane using different combinations of the samples: DES+Planck (black), DES+Planck+SUDSS (red) and DES+Planck+LSST-like (blue). We note that the fiducial cosmology is compatible with the results obtained within 1$\sigma$.}
\label{w0wa_plot}
\end{figure*}

At the top of Table \ref{table_results}, we provide our constraints for DES alone (3500 high redshift SNe Ia and 300 low redshift SNe Ia). We provide these constraints as a reference for subsequent constraints to show the likely relative improvement over forthcoming SN samples. Unfortunately, we can not directly compare these DES-only constraints to \cite{bernstein2012} as they did not provide $w$CDM predictions. However, we can compare our DES+Planck $w_{0}-w_{a}$ constraints, given in Table \ref{table_results_w0wa} and Figure \ref{w0wa_plot}, to similar predictions in \cite{bernstein2012}, and find good agreement given the different assumptions and methodologies (see figure 29 of \citealt{bernstein2012} for comparison). This provides confidence that our constraints are reasonable.

Before we present our results, we tested the stability of our methodology using 100 realisations of the same DES mock and examined the distribution of best fit values obtained using our likelihood code. As expected, the mean of the $\Omega_m$ and $w$ distributions were consistent with the fiducial cosmology (within 10\% of $\sigma$), while the mean of the distribution of fitted errors agreed with the width of the best fit distributions, and was close to the best fit errors quoted in the table from a single realisation. The spread in the error distributions of these cosmological parameters suggests there is an additional uncertainty of only a few percent on any individual realisation.

The $w$CDM constraints for our LSST-like sample of SLSNe are impressive for $\sigma_{\rm err}=0.15$. Table 1 shows that LSST alone could constrain $\Omega_m$ and $w$ to $3\%$ and $7\%$ respectively using just SLSNe. These quoted errors are only statistical  and do not account for possible differences in the absolute magnitude of SLSNe with redshift. To test such a systematic uncertainty, we have re-run the LSST-like sample in Table 1 again but marginalising over possible magnitude offsets in three bins of redshift ($z<0.8$, $0.8<z<1.3$ and $z>1.3$) using the same methodology described in Section 4 (namely allowing for a different value of $\xi$ between the three bins in equation 8). These three bins were selected to coincide with the expected ranges of redshift where major features in the spectral energy distribution of SLSNe passes through the observed filters, thus representing possible larger uncertainties in the $k$-corrections of SLSNe. In this case, the errors on $\Omega_m$ and $w$ increase to $6\%$ and $9.5\%$ respectively for the LSST-like sample alone.

We next consider in Table \ref{table_results} the likely gains in cosmological constraints from samples of SLSNe (SUDSS and LSST-like) for two possible values of the population scatter ($\sigma_{\rm err}$). For $\sigma_{\rm err}=0.15$ mag, we see that SUDSS-alone can deliver competitive cosmological constraints especially when combined with DES; the cosmological constraint on $w$ improves by $\simeq20\%$ compared to DES-alone with just 98 (73 SUDSS + 25 low-z) SLSNe. Even SUDSS on its own is competitive, close to the DES-only predictions for $\Omega_m$. Unfortunately, for the higher value of $\sigma_{\rm err}$, the extra constraining power is lost and SUDSS would likely add little to existing samples or knowledge under the assumption of $w$CDM. This is understandable given this well-defined cosmological model where dark energy becomes less important at high redshift ($z>1$). 

For the DES+LSST sample, we find constraints on $\Omega_m$ and $w$ of 2\% and 4\% respectively which are significantly better than present day errors on these parameters. We note that we have not included CMB data in these constraints as we wish to see the power of SNe alone. For comparison, we calculate the DES+Planck constraints on $\Omega_m$ and $w$ using MCMC and find constraints of 2\% on each of these parameters. Unsurprisingly the high redshift CMB measurement greatly improves the constraints on this restrictive model (constant $w$) but we stress that DES+LSST alone (in Table \ref{table_results}) delivers the same level of constraining power i.e. the errors on $\Omega_m$ and $w$ do not decrease significantly when we study DES+LSST+Planck for the $w$CDM model. We still obtain good constraints on $\Omega_m$ and $w$ from our LSST-like sample with $\sigma_{err}=0.25$.

In Table \ref{table_results_w0wa} we show the results when fitting the $w_z$CDM model (via MCMC) for different combinations of the data sets used so far. Although the level of accuracy on $\Omega_m$ is almost the same (approximately 3\%) for all the cases shown (with or without the inclusion of SLSNe), we do see a $\simeq10\%$ improvement in $w_0$ and $w_a$ when adding SUDSS (with $\sigma_{\rm err}=0.15$) to DES+Planck (Figure \ref{w0wa_plot}). This is impressive given the relatively small number of supernova added (98) and these data are being collected now. There is still some gain in constraining power for the case of $\sigma_{\rm err}=0.25$. 

The most impressive constraints come from our LSST-like sample (see Figure \ref{w0wa_plot}), combined with DES SNe Ia and Planck data. Assuming $\sigma_{\rm err}=0.15$, this combination (LSST-like+DES+Planck) should provide constraints of only 5\% and 0.143 respectively for $\left|\frac{\Delta w_0}{w_0}\right|$ and $\Delta w_a$ (see Table 2), which is competitive with constraints of 5\% and 0.16 coming from Euclid \citep[table 1.11,][]{amendola_euclid}. Therefore, SLSNe from LSST could provide a ``Stage 4" measurement of cosmology \citep{albrecht2009} when combined with the ``Stage 3" DES SN Survey. Even the lower quality LSST-like SLSNe ($\sigma_{\rm err}=0.25$) can deliver impressive dark energy constraints when combined with DES SNe Ia and Planck (see Table 2). 

\section{Conclusions}
\label{conclusions}

In this work, we have constructed realistic mock samples for Type Ia Supernovae from DES, and Superluminous Supernovae from SUDSS and possibly LSST. These mock samples are created to include the most likely sources of uncertainty when observing supernovae at cosmological distances (e.g., gravitational lensing); we marginalise over possible unknown magnitude offsets for each supernova sample (see Section 4). Tests of our methodology have shown that our predicted errors on cosmological parameters are consistent with those in the literature (e.g. \citealt{bernstein2012,betoule2014}). 

We fit these mock SN samples with cosmological models to derive likely errors on the cosmological parameters. Using our own likelihood code, we find that the addition of only 73 SLSNe expected from SUDSS (plus 25 low redshift SLSNe from other on-going surveys) to DES will improve the constraints on the equation-of-state of dark energy ($w$) and the matter density of the Universe ($\Omega_{m}$) by 20\%, assuming a flat, $w$CDM Universe and a scatter of $\sigma_{\rm err}=0.15$ mag for SLSNe (see Table 1). These data will likely be available in the next few years, leading to a significant improvement on our understanding of dark energy this decade.

We have also studied the combination of SLSNe from LSST with SNe Ia from DES. For the flat, $w$CDM model, we show that the combination of these data will deliver impressive constraints on $\Omega_{m}$ and $w$ of 2\% and 4\% respectively (LSST with statistical errors only, see Table 1). However, the real power of the LSST-like SLSNe becomes evident when we allow for a non-zero time derivative of $w=w(a)$, giving possible uncertainties of only 2\%, 5\% and 0.14 on $\Omega_{m}$, $w_0$ and $w_a$ respectively when combined with DES SNe Ia and Planck data. These errors are competitive with the predicted Euclid constraints, demonstrating a future role for Superluminous Supernonovae for probing the high redshift Universe \citep{king2014}, especially as planned forthcoming surveys like LSST should find these events in significant numbers (10,000 to $z\sim3$).

\section*{Acknowledgements}
The authors thank Gary Burton, Thomas Collett, Cosimo Inserra and Stephen Smartt for useful comments.
The author also thank the anonymous referee for the useful comments.
DS thanks the Faculty of Technology of the University of Portsmouth for support during his PhD studies.
RN and DB acknowledge support from STFC via grant ST/K00090X/1.
MS acknowledges support from the Royal Society and EU/FP7-ERC grant no [615929].

\bibliographystyle{mn2e}

\bibliography{Bibliography}

\begin{thebibliography}{}

\bibitem[\protect\citeauthoryear{{Albrecht}, {Amendola}, {Bernstein}, {Clowe},
  {Eisenstein}, {Guzzo}, {Hirata}, {Huterer}, {Kirshner}, {Kolb} \&
  {Nichol}}{{Albrecht} et~al.}{2009}]{albrecht2009}
{Albrecht} A.,  {Amendola} L.,  {Bernstein} G.,  {Clowe} D.,  {Eisenstein} D.,
  {Guzzo} L.,  {Hirata} C.,  {Huterer} D.,  {Kirshner} R.,  {Kolb} E.,
  {Nichol} R.,  2009, ArXiv e-prints

\bibitem[\protect\citeauthoryear{{Amendola}, {Appleby}, {Bacon}, {Baker},
  {Baldi}, {Bartolo}, {Blanchard}, {Bonvin}, {Borgani}, {Branchini}, {Burrage}
  \& {Camera}}{{Amendola} et~al.}{2013}]{amendola_euclid}
{Amendola} L.,  {Appleby} S.,  {Bacon} D.,  {Baker} T.,  {Baldi} M.,  {Bartolo}
  N.,  {Blanchard} A.,  {Bonvin} C.,  {Borgani} S.,  {Branchini} E.,  {Burrage}
  C.,    {Camera} S.~a.,  2013, Living Reviews in Relativity, 16, 6

\bibitem[\protect\citeauthoryear{{Amendola} \& {Tsujikawa}}{{Amendola} \&
  {Tsujikawa}}{2010}]{amendola_book}
{Amendola} L.,  {Tsujikawa} S.,  2010, {Dark Energy: Theory and Observations}

\bibitem[\protect\citeauthoryear{{Astier}, {Balland}, {Brescia}, {Cappellaro},
  {Carlberg}, {Cavuoti} \& {Della Valle}}{{Astier} et~al.}{2014}]{astier2014}
{Astier} P.,  {Balland} C.,  {Brescia} M.,  {Cappellaro} E.,  {Carlberg} R.~G.,
   {Cavuoti} S.,    {Della Valle} 2014, \aap, 572, A80

\bibitem[\protect\citeauthoryear{{Bartelmann} \& {Schneider}}{{Bartelmann} \&
  {Schneider}}{2001}]{bartelmann2001}
{Bartelmann} M.,  {Schneider} P.,  2001, \physrep, 340, 291

\bibitem[\protect\citeauthoryear{{Bernstein}, {Kessler}, {Kuhlmann}, {Biswas},
  {Kovacs}, {Aldering}, {Crane}, {D'Andrea}, {Finley} \& {Frieman}}{{Bernstein}
  et~al.}{2012}]{bernstein2012}
{Bernstein} J.~P.,  {Kessler} R.,  {Kuhlmann} S.,  {Biswas} R.,  {Kovacs} E.,
  {Aldering} G.,  {Crane} I.,  {D'Andrea} C.~B.,  {Finley} D.~A.,    {Frieman}
  2012, \apj, 753, 152

\bibitem[\protect\citeauthoryear{{Betoule}, {Kessler}, {Guy}, {Mosher},
  {Hardin}, {Biswas}, {Astier} \& {El-Hage}}{{Betoule}
  et~al.}{2014}]{betoule2014}
{Betoule} M.,  {Kessler} R.,  {Guy} J.,  {Mosher} J.,  {Hardin} D.,  {Biswas}
  R.,  {Astier} P.,    {El-Hage} 2014, \aap, 568, A22

\bibitem[\protect\citeauthoryear{{Bridle}, {Crittenden}, {Melchiorri},
  {Hobson}, {Kneissl} \& {Lasenby}}{{Bridle} et~al.}{2002}]{bridle2002}
{Bridle} S.~L.,  {Crittenden} R.,  {Melchiorri} A.,  {Hobson} M.~P.,  {Kneissl}
  R.,    {Lasenby} A.~N.,  2002, \mnras, 335, 1193

\bibitem[\protect\citeauthoryear{{Carroll}, {Press} \& {Turner}}{{Carroll}
  et~al.}{1992}]{carroll1992}
{Carroll} S.~M.,  {Press} W.~H.,    {Turner} E.~L.,  1992, \araa, 30, 499

\bibitem[\protect\citeauthoryear{{Chevallier} \& {Polarski}}{{Chevallier} \&
  {Polarski}}{2001}]{chevallier2001}
{Chevallier} M.,  {Polarski} D.,  2001, International Journal of Modern Physics
  D, 10, 213

\bibitem[\protect\citeauthoryear{{Conley}, {Guy}, {Sullivan}, {Regnault},
  {Astier}, {Balland}, {Basa}, {Carlberg}, {Fouchez}, {Hardin} \&
  {Hook}.}{{Conley} et~al.}{2011}]{conley2011}
{Conley} A.,  {Guy} J.,  {Sullivan} M.,  {Regnault} N.,  {Astier} P.,
  {Balland} C.,  {Basa} S.,  {Carlberg} R.~G.,  {Fouchez} D.,  {Hardin} D.,
  {Hook}. 2011, \apjs, 192, 1

\bibitem[\protect\citeauthoryear{{Cooke}, {Sullivan}, {Gal-Yam}, {Barton},
  {Carlberg}, {Ryan-Weber}, {Horst}, {Omori} \& {D{\'{\i}}az}}{{Cooke}
  et~al.}{2012}]{cooke2012}
{Cooke} J.,  {Sullivan} M.,  {Gal-Yam} A.,  {Barton} E.~J.,  {Carlberg} R.~G.,
  {Ryan-Weber} E.~V.,  {Horst} C.,  {Omori} Y.,    {D{\'{\i}}az} C.~G.,  2012,
  \nat, 491, 228

\bibitem[\protect\citeauthoryear{{Dodelson} \& {Vallinotto}}{{Dodelson} \&
  {Vallinotto}}{2006}]{dodelson2005}
{Dodelson} S.,  {Vallinotto} A.,  2006, \prd, 74, 063515

\bibitem[\protect\citeauthoryear{{Eisenstein} \& {Hu}}{{Eisenstein} \&
  {Hu}}{1999}]{eisenstein1999}
{Eisenstein} D.~J.,  {Hu} W.,  1999, \apj, 511, 5

\bibitem[\protect\citeauthoryear{{Flaugher}, {Diehl}, {Honscheid}, {Abbott},
  {Alvarez}, {Angstadt}, {Annis}, {Antonik}, {Ballester}, {Beaufore},
  {Bernstein}, {Bernstein}, {Bigelow}, {Bonati} \& {Boprie}}{{Flaugher}
  et~al.}{2015}]{flaugher2015}
{Flaugher} B.,  {Diehl} H.~T.,  {Honscheid} K.,  {Abbott} T.~M.~C.,  {Alvarez}
  O.,  {Angstadt} R.,  {Annis} J.~T.,  {Antonik} M.,  {Ballester} O.,
  {Beaufore} L.,  {Bernstein} G.~M.,  {Bernstein} R.~A.,  {Bigelow} B.,
  {Bonati} M.,    {Boprie} 2015, \aj, 150, 150

\bibitem[\protect\citeauthoryear{{Gal-Yam}}{{Gal-Yam}}{2012}]{Gal-Yam2012}
{Gal-Yam} A.,  2012, Science, 337, 927

\bibitem[\protect\citeauthoryear{{Garnavich}, {Kirshner}, {Challis}, {Tonry},
  {Gilliland}, {Smith} \& {Clocchiatti}}{{Garnavich}
  et~al.}{1998}]{garnavich1997}
{Garnavich} P.~M.,  {Kirshner} R.~P.,  {Challis} P.,  {Tonry} J.,  {Gilliland}
  R.~L.,  {Smith} R.~C.,    {Clocchiatti} 1998, \apjl, 493, L53

\bibitem[\protect\citeauthoryear{{Ghirlanda}, {Ghisellini} \&
  {Firmani}}{{Ghirlanda} et~al.}{2006}]{ghirlanda2006}
{Ghirlanda} G.,  {Ghisellini} G.,    {Firmani} C.,  2006, New Journal of
  Physics, 8, 123

\bibitem[\protect\citeauthoryear{{Howell}, {Kasen}, {Lidman}, {Sullivan},
  {Conley}, {Astier}, {Balland}, {Carlberg}, {Fouchez}, {Guy}, {Hardin} \&
  {Pain}}{{Howell} et~al.}{2013}]{howell2013}
{Howell} D.~A.,  {Kasen} D.,  {Lidman} C.,  {Sullivan} M.,  {Conley} A.,
  {Astier} P.,  {Balland} C.,  {Carlberg} R.~G.,  {Fouchez} D.,  {Guy} J.,
  {Hardin} D.,    {Pain} 2013, \apj, 779, 98

\bibitem[\protect\citeauthoryear{{Inserra} \& {Smartt}}{{Inserra} \&
  {Smartt}}{2014}]{inserra2014}
{Inserra} C.,  {Smartt} S.~J.,  2014, \apj, 796, 87

\bibitem[\protect\citeauthoryear{{J{\"o}nsson}, {Sullivan}, {Hook}, {Basa},
  {Carlberg}, {Conley}, {Fouchez}, {Howell}, {Perrett} \&
  {Pritchet}}{{J{\"o}nsson} et~al.}{2010}]{jonsson2010}
{J{\"o}nsson} J.,  {Sullivan} M.,  {Hook} I.,  {Basa} S.,  {Carlberg} R.,
  {Conley} A.,  {Fouchez} D.,  {Howell} D.~A.,  {Perrett} K.,    {Pritchet} C.,
   2010, \mnras, 405, 535

\bibitem[\protect\citeauthoryear{{King}, {Davis}, {Denney}, {Vestergaard} \&
  {Watson}}{{King} et~al.}{2014}]{king2014}
{King} A.~L.,  {Davis} T.~M.,  {Denney} K.~D.,  {Vestergaard} M.,    {Watson}
  D.,  2014, \mnras, 441, 3454

\bibitem[\protect\citeauthoryear{{Lewis} \& {Bridle}}{{Lewis} \&
  {Bridle}}{2002}]{cosmomc}
{Lewis} A.,  {Bridle} S.,  2002, \prd, 66, 103511

\bibitem[\protect\citeauthoryear{{LSST Science Collaboration}, {Abell},
  {Allison}, {Anderson}, {Andrew}, {Angel}, {Armus}, {Arnett}, {Asztalos},
  {Axelrod} \& et al.}{{LSST Science Collaboration}
  et~al.}{2009}]{lsst_science_book}
{LSST Science Collaboration} {Abell} P.~A.,  {Allison} J.,  {Anderson} S.~F.,
  {Andrew} J.~R.,  {Angel} J.~R.~P.,  {Armus} L.,  {Arnett} D.,  {Asztalos}
  S.~J.,  {Axelrod} T.~S.,    et al. 2009, ArXiv e-prints

\bibitem[\protect\citeauthoryear{{Marra}, {Quartin} \& {Amendola}}{{Marra}
  et~al.}{2013}]{marra2013}
{Marra} V.,  {Quartin} M.,    {Amendola} L.,  2013, \prd, 88, 063004

\bibitem[\protect\citeauthoryear{{Nicholl}, {Smartt}, {Jerkstrand}, {Inserra},
  {McCrum}, {Kotak}, {Fraser}, {Wright}, {Chen}, {Smith}, {Young}, {Sim},
  {Valenti}, {Howell} \& {Bresolin}}{{Nicholl} et~al.}{2013}]{nicholl2013}
{Nicholl} M.,  {Smartt} S.~J.,  {Jerkstrand} A.,  {Inserra} C.,  {McCrum} M.,
  {Kotak} R.,  {Fraser} M.,  {Wright} D.,  {Chen} T.-W.,  {Smith} K.,  {Young}
  D.~R.,  {Sim} S.~A.,  {Valenti} S.,  {Howell} D.~A.,    {Bresolin} 2013,
  \nat, 502, 346

\bibitem[\protect\citeauthoryear{{Papadopoulos}, {D'Andrea}, {Sullivan},
  {Nichol}, {Barbary}, {Biswas}, {Brown}, {Covarrubias}, {Finley} \&
  {Fischer}}{{Papadopoulos} et~al.}{2015}]{Papadopoulos2015}
{Papadopoulos} A.,  {D'Andrea} C.~B.,  {Sullivan} M.,  {Nichol} R.~C.,
  {Barbary} K.,  {Biswas} R.,  {Brown} P.~J.,  {Covarrubias} R.~A.,  {Finley}
  D.~A.,    {Fischer} 2015, \mnras, 449, 1215

\bibitem[\protect\citeauthoryear{{Perlmutter}, {Aldering}, {Goldhaber}, {Knop},
  {Nugent}, {Castro}, {Deustua}, {Fabbro}, {Goobar}, {Groom} \&
  {Hook}}{{Perlmutter} et~al.}{1999}]{perl99}
{Perlmutter} S.,  {Aldering} G.,  {Goldhaber} G.,  {Knop} R.~A.,  {Nugent} P.,
  {Castro} P.~G.,  {Deustua} S.,  {Fabbro} S.,  {Goobar} A.,  {Groom} D.~E.,
  {Hook} 1999, \apj, 517, 565

\bibitem[\protect\citeauthoryear{{Planck Collaboration}, {Ade}, {Aghanim},
  {Armitage-Caplan}, {Arnaud}, {Ashdown}, {Atrio-Barandela}, {Aumont},
  {Baccigalupi}, {Banday} \& et al.}{{Planck Collaboration}
  et~al.}{2014}]{planck_collaboration}
{Planck Collaboration} {Ade} P.~A.~R.,  {Aghanim} N.,  {Armitage-Caplan} C.,
  {Arnaud} M.,  {Ashdown} M.,  {Atrio-Barandela} F.,  {Aumont} J.,
  {Baccigalupi} C.,  {Banday} A.~J.,    et al. 2014, \aap, 571, A16

\bibitem[\protect\citeauthoryear{{Press}, {Teukolsky}, {Vetterling} \&
  {Flannery}}{{Press} et~al.}{1992}]{press1992}
{Press} W.~H.,  {Teukolsky} S.~A.,  {Vetterling} W.~T.,    {Flannery} B.~P.,
  1992, {Numerical recipes in FORTRAN. The art of scientific computing}

\bibitem[\protect\citeauthoryear{{Quartin}, {Marra} \& {Amendola}}{{Quartin}
  et~al.}{2014}]{quartin2014}
{Quartin} M.,  {Marra} V.,    {Amendola} L.,  2014, \prd, 89, 023009

\bibitem[\protect\citeauthoryear{{Quimby}, {Yuan}, {Akerlof}, {Wheeler} \&
  {Warren}}{{Quimby} et~al.}{2012}]{quimby2012}
{Quimby} R.~M.,  {Yuan} F.,  {Akerlof} C.,  {Wheeler} J.~C.,    {Warren} M.~S.,
   2012, \aj, 144, 177

\bibitem[\protect\citeauthoryear{{Rau}, {Kulkarni}, {Law}, {Bloom}, {Ciardi},
  {Djorgovski} \& {Fox}}{{Rau} et~al.}{2009}]{ptf}
{Rau} A.,  {Kulkarni} S.~R.,  {Law} N.~M.,  {Bloom} J.~S.,  {Ciardi} D.,
  {Djorgovski} G.~S.,    {Fox} 2009, \pasp, 121, 1334

\bibitem[\protect\citeauthoryear{{Riess}, {Filippenko}, {Challis},
  {Clocchiatti}, {Diercks} \& {Garnavich}}{{Riess} et~al.}{1998}]{riess98}
{Riess} A.~G.,  {Filippenko} A.~V.,  {Challis} P.,  {Clocchiatti} A.,
  {Diercks} A.,    {Garnavich} 1998, \aj, 116, 1009

\bibitem[\protect\citeauthoryear{{Schmidt}}{{Schmidt}}{2012}]{skymapper}
{Schmidt} B.~P.,  2012, in American Astronomical Society Meeting Abstracts 220
  Vol.~220 of American Astronomical Society Meeting Abstracts, {SkyMapper:
  Surveying the Southern Sky}.
p. 426.01

\bibitem[\protect\citeauthoryear{Schneider, Meylan, Kochanek, Jetzer, North \&
  Wambsganss}{Schneider et~al.}{2006}]{schneider2005}
Schneider P.,  Meylan G.,  Kochanek C.,  Jetzer P.,  North P.,    Wambsganss
  J.,  2006, Gravitational Lensing: Strong, Weak and Micro: Saas-Fee Advanced
  Course 33.
Saas-Fee Advanced Course, Springer Berlin Heidelberg

\bibitem[\protect\citeauthoryear{{Scoville}, {Abraham}, {Aussel}, {Barnes},
  {Benson}, {Blain}, {Calzetti}, {Comastri}, {Capak}, {Carilli}, {Carlstrom} \&
  {Carollo}}{{Scoville} et~al.}{2007}]{scoville2007}
{Scoville} N.,  {Abraham} R.~G.,  {Aussel} H.,  {Barnes} J.~E.,  {Benson} A.,
  {Blain} A.~W.,  {Calzetti} D.,  {Comastri} A.,  {Capak} P.,  {Carilli} C.,
  {Carlstrom} J.~E.,    {Carollo} 2007, \apjs, 172, 38

\bibitem[\protect\citeauthoryear{{Smith}, {Bacon}, {Nichol}, {Campbell},
  {Clarkson}, {Maartens}, {D'Andrea}, {Bassett}, {Cinabro}, {Finley},
  {Frieman}, {Galbany}, {Garnavich}, {Olmstead}, {Schneider}, {Shapiro} \&
  {Sollerman}}{{Smith} et~al.}{2014}]{smith2014}
{Smith} M.,  {Bacon} D.~J.,  {Nichol} R.~C.,  {Campbell} H.,  {Clarkson} C.,
  {Maartens} R.,  {D'Andrea} C.~B.,  {Bassett} B.~A.,  {Cinabro} D.,  {Finley}
  D.~A.,  {Frieman} J.~A.,  {Galbany} L.,  {Garnavich} P.~M.,  {Olmstead}
  M.~D.,  {Schneider} D.~P.,  {Shapiro} C.,    {Sollerman} J.,  2014, \apj,
  780, 24

\bibitem[\protect\citeauthoryear{{Smith}, {Peacock}, {Jenkins}, {White},
  {Frenk}, {Pearce}, {Thomas}, {Efstathiou} \& {Couchman}}{{Smith}
  et~al.}{2003}]{smith2003}
{Smith} R.~E.,  {Peacock} J.~A.,  {Jenkins} A.,  {White} S.~D.~M.,  {Frenk}
  C.~S.,  {Pearce} F.~R.,  {Thomas} P.~A.,  {Efstathiou} G.,    {Couchman}
  H.~M.~P.,  2003, \mnras, 341, 1311

\bibitem[\protect\citeauthoryear{{Suzuki}, {Rubin}, {Lidman}, {Aldering},
  {Amanullah}, {Barbary}, {Barrientos}, {Botyanszki}, {Brodwin}, {Connolly} \&
  {Dawson}}{{Suzuki} et~al.}{2012}]{union21}
{Suzuki} N.,  {Rubin} D.,  {Lidman} C.,  {Aldering} G.,  {Amanullah} R.,
  {Barbary} K.,  {Barrientos} L.~F.,  {Botyanszki} J.,  {Brodwin} M.,
  {Connolly} N.,    {Dawson} 2012, \apj, 746, 85

\bibitem[\protect\citeauthoryear{{Teukolsky}, {Vetterling}, {Flannery}, {Lloyd}
  \& {Rees}}{{Teukolsky} et~al.}{1993}]{press1993}
{Teukolsky} S.~A.,  {Vetterling} W.~T.,  {Flannery} B.~P.,  {Lloyd} C.,
  {Rees} P.,  1993, The Observatory, 113, 214

\bibitem[\protect\citeauthoryear{{Watson}, {Denney}, {Vestergaard} \&
  {Davis}}{{Watson} et~al.}{2011}]{watson2011}
{Watson} D.,  {Denney} K.~D.,  {Vestergaard} M.,    {Davis} T.~M.,  2011,
  \apjl, 740, L49

\bibitem[\protect\citeauthoryear{{Wei}, {Wu} \& {Melia}}{{Wei}
  et~al.}{2015}]{wei2015}
{Wei} J.-J.,  {Wu} X.-F.,    {Melia} F.,  2015, \aj, 149, 165

\bibitem[\protect\citeauthoryear{{Young}, {Smartt}, {Mattila}, {Tanvir},
  {Bersier}, {Chambers}, {Kaiser} \& {Tonry}}{{Young} et~al.}{2008}]{panSTARRS}
{Young} D.~R.,  {Smartt} S.~J.,  {Mattila} S.,  {Tanvir} N.~R.,  {Bersier} D.,
  {Chambers} K.~C.,  {Kaiser} N.,    {Tonry} J.~L.,  2008, \aap, 489, 359

\end{thebibliography}

\end{document}